\documentclass[twocolumn,aps,prl,superscriptaddress,amsfonts,showpacs]{revtex4-1}

\usepackage{amsmath}
\usepackage{amssymb}
\usepackage{bm}
\usepackage{graphicx}
\usepackage{etoolbox}
\usepackage{siunitx}

\newcommand{\spup}{\ensuremath{ \!\!\uparrow }}
\newcommand{\spdn}{\ensuremath{ \!\!\downarrow}}

\newcommand{\kb}{\ensuremath{k_{\mathrm{B}}}}
\newcommand{\nspup}{ \makebox[0.7em][c]{$\uparrow$}}
\newcommand{\nspdn}{ \makebox[0.7em][c]{$\downarrow$}}

\usepackage{soul,xcolor}

\newcommand{\FigureOne}{
\begin{figure}
\centering\includegraphics[width=\columnwidth]{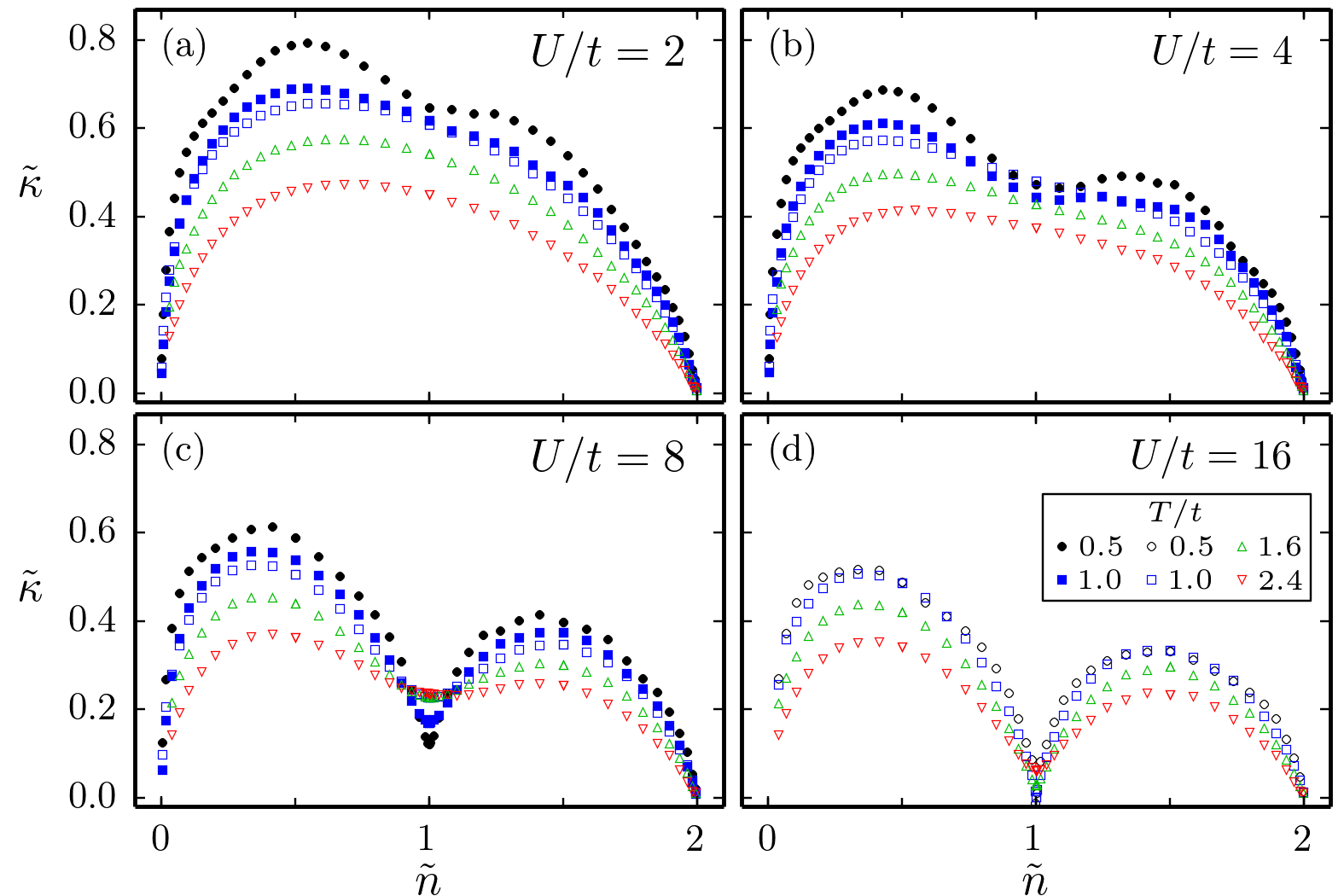} \caption{(color
online) Normalized compressibility versus density for the homogeneous 3D
Hubbard model, shown for various interaction strengths and temperatures.  The
different curves were obtained using DQMC (closed symbols) and NLCE (open
symbols).  At half-filling, $\tilde{n}=1$, the compressibility vanishes for
strong interactions and low temperatures as the system enters the Mott
insulating regime.  }
\end{figure}
}

\newcommand{\FigureTwo}{
\begin{figure}
\centering\includegraphics[width=\columnwidth]{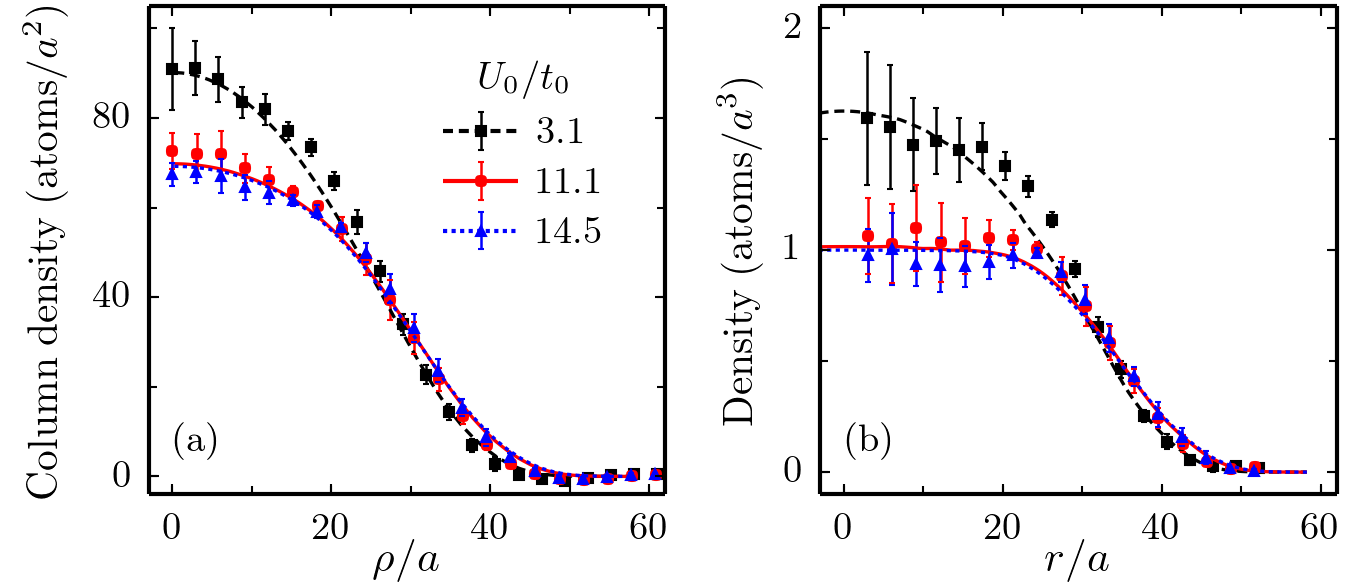} \caption{(color
online) (a) Azimuthally averaged column density (including both spin states)
vs. distance from the imaging axis $\rho$, for different values of
$U_{0}/t_{0}$.  Data points represent the average of eight individual
realizations, with error bars corresponding to the standard deviation.  The
lines in (a) are obtained by integrating the density (calculated for
$N=2\times10^{5}$ atoms at $T/t_{0}=0.6$) along the imaging axis. (b) Data
points correspond to density profiles extracted from the column densities using
the inverse Abel transform, where $r$ is the distance from the center of the
trap.  The lines in (b) show the density calculated for our trap along a body
diagonal of the lattice.  }	
\end{figure}
}

\newcommand{\FigureThree}{
\begin{figure}
\centering\includegraphics[width=\columnwidth]{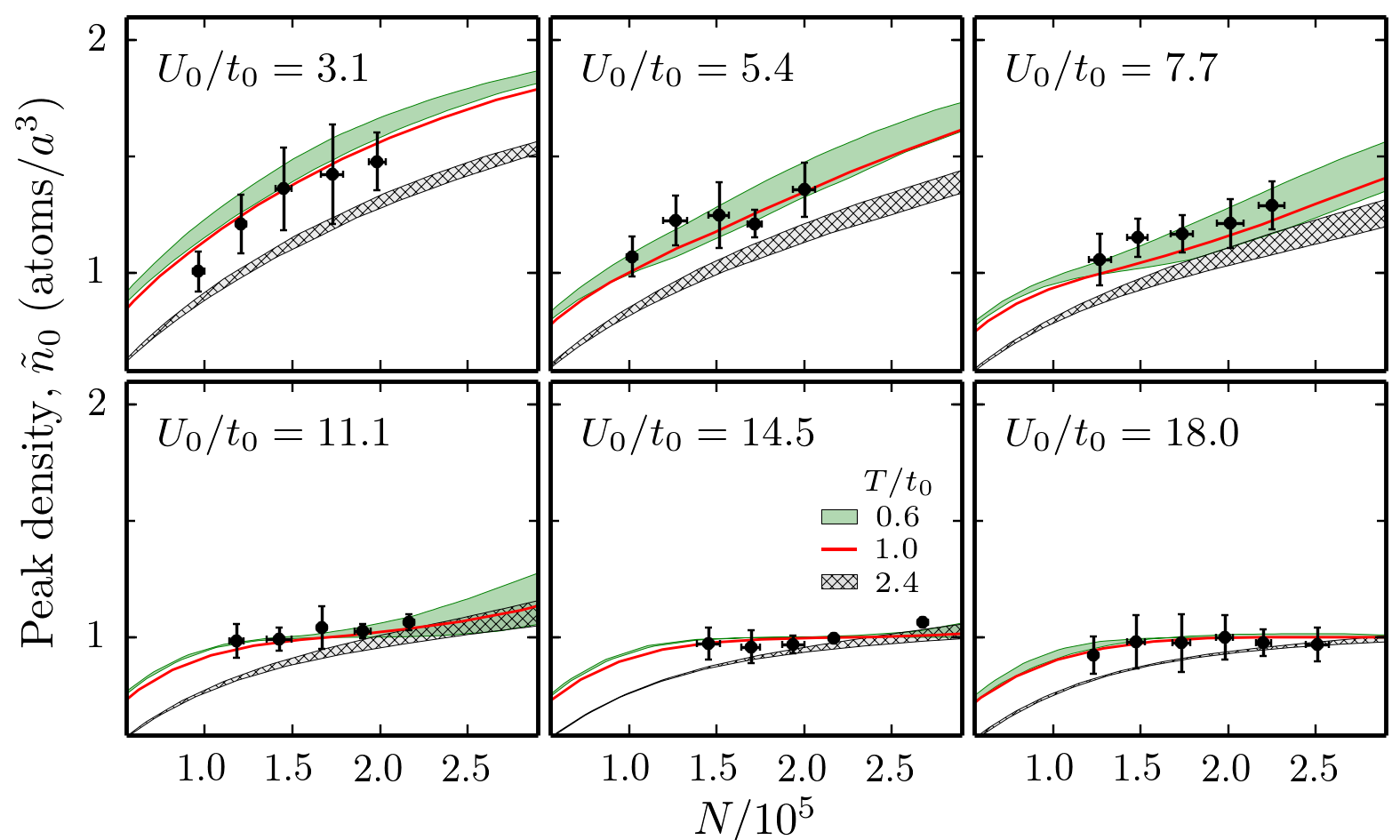} \caption{ (color
online) Central density, $\tilde{n}_{0}$ vs. atom number for various
interaction strengths.  The symbols show the average for a set of 5 to 10
independent realizations, with error bars indicating the standard deviation.
The shaded regions are the results of numerical calculations for our trap at
$T/t_{0}=0.6$ (solid, green) and 2.4 (crosshatched, gray), with the width of
each region corresponding to a $\pm$14\% systematic uncertainty in the value of
$U_{0}/t_{0}$, arising from the $\pm$5\% uncertainty in $V_{0}$. The red line
is calculated at $T/t_{0}=1.0$, without considering the trap systematics. The
calculated density becomes relatively insensitive to uncertainties in
$U_{0}/t_{0}$ for the two larger values of $U_{0}/t_{0}$, which are deep in the
Mott regime.  For $T/t_{0}=0.6$ the total entropy per particle, $S/(N\kb)$, is
between 0.5 and 1.0 for the ranges of $N$ and $U_{0}/t_{0}$ shown in the
figure.  A temperature of $T/t_{0}=2.4$ is chosen for comparison, as in this
case $S/(N\kb)$ is between 1.5 and 2.4, which is similar to the range between
1.6 and 2.2 reported from the analysis of a previous
experiment~\cite{Jordens2010}.}
\end{figure}
}

\newcommand{\FigureFour}{
\begin{figure}
\centering\includegraphics[width=\columnwidth]{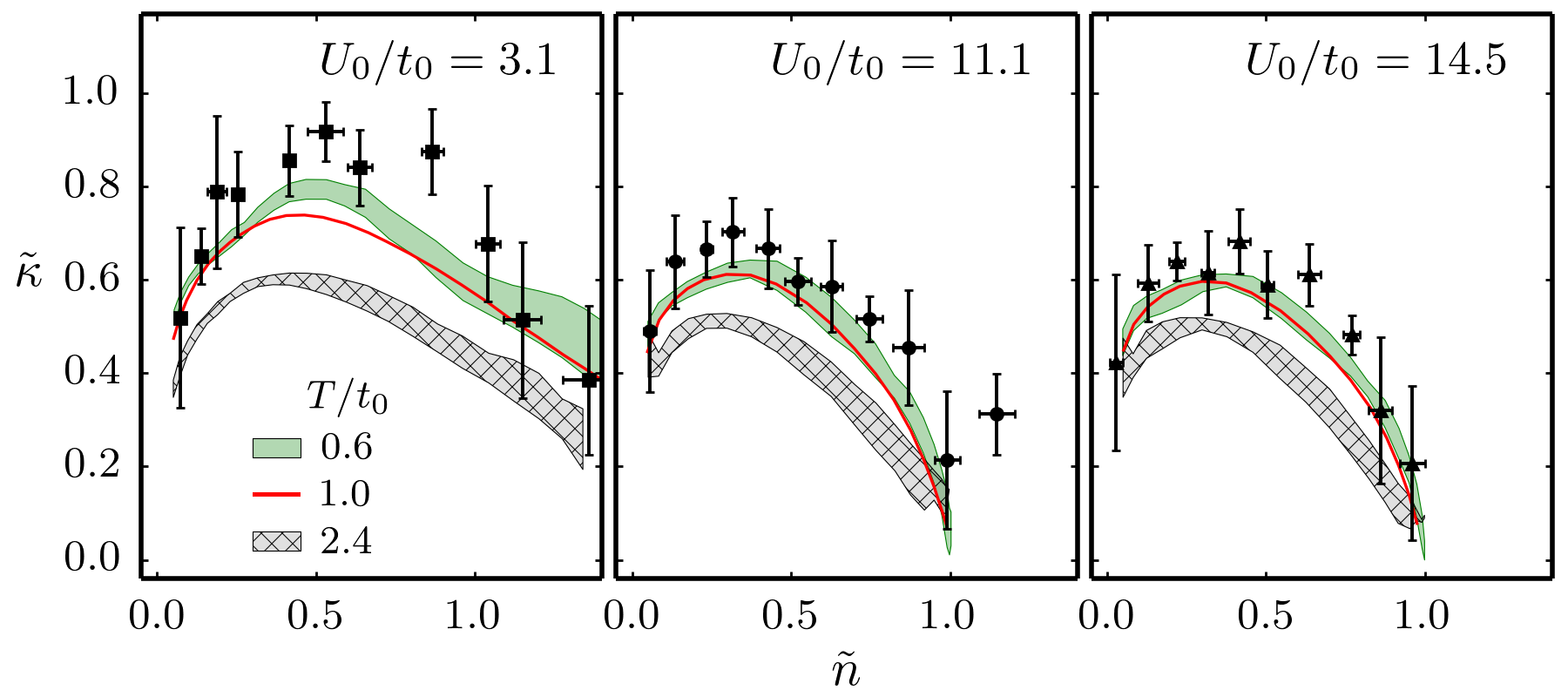} \caption{(color
online) Normalized local compressibility, $\tilde{\kappa}$ versus density for
different values of $U_{0}/t_{0}$.  Closed symbols show the average of eight
individual realizations with error bars indicating the standard deviation.  The
shaded regions are numerical calculations at $T/t_{0}=0.6$ (solid, green)  and
$T/t_{0}=2.4$ (crosshatched, gray) for $N=2\times 10^{5}$, where the width of
the region reflects a $\pm$14\% systematic uncertainty in $U_{0}/t_{0}$.
The red line is calculated at $T/t_{0}=1.0$, without considering the trap
systematics.  With $N=2\times 10^{5}$, the total entropy per particle at
$T/t_{0}=0.6$ is approximately $S/(N\kb)=0.58$, 0.76, and 0.82 for $U_{0}/t_{0}
= 3.1,\, 11.1,\ \mathrm{and}\ 14.5$, respectively; at $T/t_{0}=2.4$ it is
approximately $S/(N\kb)=1.59$, 1.70, and 1.66, respectively.  }
\end{figure}
}

\emergencystretch=\maxdimen
\begin{document}

\title{Compressibility of a fermionic Mott insulator of ultracold atoms}

\author{Pedro M. Duarte}
\affiliation{Department of Physics and Astronomy and Rice Quantum Institute, Rice
University, Houston, TX 77005, USA}
\author{Russell A. Hart}
\affiliation{Department of Physics and Astronomy and Rice Quantum Institute, Rice
University, Houston, TX 77005, USA}

\author{Tsung-Lin Yang}
\author{Xinxing Liu} 
\affiliation{Department of Physics and Astronomy and Rice Quantum Institute, Rice
University, Houston, TX 77005, USA} 

\author{Thereza Paiva}
\affiliation{Instituto de Fisica, Universidade Federal do Rio de Janeiro Cx.P.
68.528, 21941-972 Rio de Janeiro RJ, Brazil}

\author{Ehsan Khatami}
\affiliation{Department of Physics and Astronomy, San Jose State University, San
Jose, CA 95192, USA}

\author{Richard T. Scalettar} 
\affiliation{Department of Physics, University of California, Davis, California
95616, USA}

\author{Nandini Trivedi}
\affiliation{Department of Physics, The Ohio State University, Columbus, Ohio 43210,
USA}

\author{Randall G. Hulet}
\email{randy@rice.edu} 
\affiliation{Department of Physics and Astronomy and Rice Quantum Institute, Rice
University, Houston, TX 77005, USA}

\date{\today}

\begin{abstract}
We characterize the Mott insulating regime of a repulsively interacting Fermi
gas of ultracold atoms in a three-dimensional optical lattice.  We use {\it
in-situ} imaging to extract the central density of the gas, and to determine
its local compressibility.  For intermediate to strong interactions, we observe
the emergence of a plateau in the density as a function of atom number, and a
reduction of the compressibility at a density of one atom per site, indicating
the formation of a Mott insulator.  Comparisons to state-of-the-art numerical
simulations of the Hubbard model over a wide range of interactions reveal that
the temperature of the gas is of the order of, or below, the tunneling energy
scale.  Our results hold great promise for the exploration of many-body
phenomena with ultracold atoms, where the local compressibility can be a useful
tool to detect signatures of different phases or phase boundaries at specific
values of the filling.
\end{abstract}

\maketitle

The Hubbard model, which  describes spin-$1/2$ fermions in a lattice with
on-site interactions, is one of the fundamental models in quantum many-body
physics.  It is a notable example of how strongly correlated phases emerge from
simple Hamiltonians: it exhibits a Mott insulating regime, antiferromagnetism,
and is widely believed to support a $d$-wave superfluid state in two dimensions
(2D),  which could explain high-temperature superconductivity as observed in
the cuprates~\cite{ANDERSON87}.  Despite intense efforts, an exact solution of
the Hubbard model in more than one dimension and for arbitrary filling has
evaded theoretical and computational approaches to this day.  Complementing
these approaches, the last decade has seen the development of ultracold atoms
in optical lattices as a new and versatile platform for the study of many-body
physics~\cite{Bloch2008,Jaksch2005}.  In this work, we study a two-spin
component degenerate gas of fermions in a simple cubic lattice, a system which
realizes the three-dimensional (3D) single band Hubbard model.  
 
Previous ground-breaking experiments investigated the Mott transition in
trapped lattice fermions by measuring the variation of the bulk double
occupancy with atom number~\cite{Jordens2008,Scarola2009,Taie2012} and the
response of the cloud radius to changes in external
confinement~\cite{Schneider2008}, both of which are related to the global
compressibility.  Several key issues, however, remain to be addressed: (i) As
bulk measurements are the result of an average over both metallic and
insulating phases simultaneously present in the trap, how does the {\it local}
compressibility behave within the trap?  (ii)  How does the compressibility
respond at lower temperatures, as one approaches the magnetic transition?
(iii)  Can more robust theoretical treatments be employed to benchmark the
observed behavior?

In this paper, we address these issues, making significant progress towards
understanding the physics of the fermionic Hubbard Hamiltonian through optical
lattice emulation.  We extract the local compressibility of the gas from a
measurement of the {\it in-situ} density profile, a procedure that has been
previously demonstrated for a Fermi gas in a harmonic
potential~\cite{PhysRevA.85.063615}, and for lattice bosons~\cite{Gemelke2009}.
The local compressibility, as well as the central density of the gas, are
readily compared with numerical simulations within the local density
approximation (LDA).   Previous work has shown that the LDA agrees well with
numerical calculations of the inhomogeneous Hubbard Hamiltonian away from the
quantum critical regime close to the N\'{e}el
transition~\cite{Rigol2003,Helmes2008,Chiesa2011}.  The local character of our
measurements allows differentiation between the incompressible Mott insulating
core and the compressible surrounding metal, thus enabling a more precise
characterization of the Mott transition, even at intermediate values of
the coupling strength, where magnetic correlations are predicted to be
strongest~\cite{Staudt2000, Paiva2011, Kozik2013}.

The Hubbard Hamiltonian is given by
\begin{equation}
    \hat{H} = -t \sum_{\langle ij \rangle, \sigma } \left(  
          \hat{c}_{i\sigma}^{\dagger} \hat{c}_{j\sigma}^{\phantom{\dagger}} 
         + \text{h.c.}  \right)
        + U \sum_{i} \hat{n}_{i\ \spup} \hat{n}_{i\ \spdn} 
        - \mu \sum_{i,\sigma} \hat{n}_{i\sigma}.
\end{equation} 
Here, the indices $i,j$ denote lattice sites, the spin states are labeled as
$\sigma=\nspup\ \text{or}\ \nspdn\,$, the angled brackets indicate summation
over nearest-neighbors,  $t$ is the nearest-neighbor tunneling matrix element,
$U$ ($>0$) is the on-site interaction energy, $\mu$ is the chemical potential,
$\hat{c}_{i\sigma}^{\dagger}$ ($\hat{c}_{i\sigma}^{\phantom{\dagger}}$) is the
creation (annihilation) operator for a fermion with spin $\sigma$ at site $i$,
and $\hat{n}_{i\sigma} = \hat{c}_{i\sigma}^{\dagger}
\hat{c}_{i\sigma}^{\phantom{\dagger}}$ is the density operator.

For $\mu=U/2$, the average density of the system is $n=1$ particle per lattice
site (half-filling).  At half-filling, as the temperature $T$ is reduced, or as
$U$ is increased, such that $T\ll U$, the system undergoes a smooth crossover
to a Mott insulating regime, characterized by a suppression of the number of
doubly occupied sites and a suppression of density fluctuations, which implies
a reduction of the compressibility~\cite{PhysRevLett.101.210403}.  If $T$ is
reduced below the N\'{e}el temperature $T_{N}$ ($\sim\!4t^{2}/U$ for $U\gg t$),
the system undergoes a phase transition to an antiferromagnetic (AFM) state.

Cooling and thermometry have been the greatest challenges for realizing the
Hubbard model with ultracold atoms in optical lattices~\cite{McKay2011}.  Even
though the temperatures required for pairing and superfluidity in the doped
Hubbard model~\cite{Hofstetter2002} have not yet been reached, the past few
years have seen steady experimental progress.  This includes the observation of
Fermi surfaces in a band insulator~\cite{Kohl2005},  the observation of the
Mott insulating regime for strong couplings ($U/t\geq 18$)~\cite{Jordens2008,
Schneider2008, Taie2012} 
and, more recently, the detection of AFM spin correlations in 1D
chains~\cite{Greif2013,Imriska2014} and in a 3D lattice~\cite{Hart2014arxiv}.

A vanishing local compressibility characterizes the Mott regime in the Hubbard
model.  It can also be a useful observable to characterize other phases and
models realized with ultracold atoms. For example, kinks in the local
compressibility can indicate phase boundaries in the trapped
system~\cite{PhysRevLett.103.085701}. The isothermal compressibility of a gas
is defined as 
\begin{equation}
    \kappa = \frac{1}{n^{2}} \frac{ \partial n }{ \partial \mu }. 
\end{equation} 
For atoms in a 3D lattice we consider the unitless quantity $(t/a^{3})\kappa$,
where $a$ is the lattice spacing.  In the limit of zero lattice depth,
$t\rightarrow - \frac{a}{2\pi} \int_{-\pi/a}^{\pi/a} \frac{\hbar^{2}q^{2}}{2m}
\exp[ i q a]\,\text{d}q = (2/\pi^{2}) E_{r}$, where $q$ is the quasimomentum,
$E_{r} = \frac{ \hbar^{2} \pi^{2}}{ 2ma^{2} } $ is the recoil energy, and $m$
is the mass of the particles.   For a free Fermi gas with no interactions, the
compressibility at zero temperature is given by $\kappa_{0} = \frac{
3}{2nE_{F}}$, where $E_{F}$ is the Fermi energy for each spin component. In
this paper we consider the normalized compressibility $\tilde{\kappa}$, defined
as 
\begin{equation} 
\tilde{\kappa} \equiv  \frac{ (t/a^{3})\kappa  } 
                       {( (2\pi^{2}/E_{r})  /a^{3}) \kappa_{0}} =  
             \frac{ (3\pi^{2})^{2/3} }{ 2 } 
             \frac{ \partial \tilde{n}^{2/3} }{ \partial (\mu/t) },
\end{equation}
where $\tilde{n} = a^{3}n$.

\FigureOne

We start by presenting theoretical results for $\tilde{\kappa}$, which underlie
the interpretation of our experimental results.  In Fig.~1 we show theoretical
results for $\tilde{\kappa}$ at various values of $T/t$ and $U/t$, obtained
using determinantal quantum Monte Carlo
(DQMC)~\cite{Blankenbecler1981,Paiva2010} and a numerical linked-cluster
expansion (NLCE)~\cite{Rigol2006,Khatami2011,Note1} 
up to the eighth order in the site expansion. These two methods complement each 
other, and provide results over a wide range of interactions and temperatures.  
While NLCE can reach lower temperatures than DQMC at large $U/t$, the opposite 
is true at weak coupling.  Figure~1 shows that the theoretical compressibility 
diminishes at half-filling and larger $U/t$ as the system enters the Mott 
insulating regime, and at $\tilde{n}=2$, where a band insulator forms. In 
addition, Fig.~1 demonstrates that at a temperature $T\leq t$, locally 
resolving the compressibility enables one to observe the Mott regime for 
coupling strengths as low as $U/t\sim8$, in the vicinity of the interaction 
strength that maximizes $T_{N}$~\cite{Staudt2000,Paiva2011,Kozik2013}, rather 
than requiring larger couplings~\cite{Jordens2008, Schneider2008, Taie2012}.

In our experiment, we produce a two-spin component degenerate Fermi gas of
$^{6}$Li atoms in the $|F=1/2; m_{F}=+1/2\rangle$ and $|F=1/2;
m_{F}=-1/2\rangle$ hyperfine states, which we label $|\spup\rangle$ and
$|\spdn\rangle$, respectively. The apparatus has been described
previously~\cite{Duarte2011,Hart2014arxiv}.  Briefly, the spin mixture is
evaporated into a harmonic dimple trap and then loaded into a simple cubic
optical lattice.  We control the total number of atoms, $N$, by adjusting the
final depth of the dimple trap.  The temperature of the atoms in the dimple is
measured by fitting the density distribution after time of flight. We obtain
$T/T_{F} = 0.04\pm0.02$, independent of $N$ within the range of atom numbers
considered for this paper.

The optical lattice is formed by three retroreflected red-detuned (1064 nm)
Gaussian laser beams of depth $V_{0}=7\,E_{r}$.  The lattice depth is
calibrated via lattice phase modulation spectroscopy, up to a systematic
uncertainty of $\pm$5\%.  Due to the Gaussian beam profiles, the lattice depth
decreases with distance from the center, which results in increasing $t$ and
decreasing $U/t$.  The lattice depth varies along the 111 body diagonals as
$V(r) = V_{0} \exp[-4r^{2}/(3w_{L}^{2})]$, where $V_{0}$ is the lattice depth
at the center, $r$ is the distance from the center, and $w_{L}$ is the waist
($1/e^2$ radius) of the lattice beams. We make use of the broad Feshbach
resonance in $^{6}$Li at 832~G~\cite{Houbiers1998,Zurn2013} to set the on-site
interaction strength, $U$.

The lattice confinement is compensated by the addition of three blue-detuned
(532 nm) Gaussian beams, which overlap each of the lattice beams but are not
themselves retroreflected~\cite{Hart2014arxiv,Mathy2012}.    The overall
confinement in the lattice, which sets the density of the cloud, is adjusted by
changing the intensity of the compensation beams. We create samples which
appear spherically symmetric with slight adjustment of the intensity of the
three independent compensation beams. The average value of the compensation
depth is set at $3.8\,E_{r}$, with a systematic $\pm$10\% relative error
resulting from the calibration of $w_{L}$ and the compensation beam waists,
$w_{C}$.  The beam waists along each axis are calibrated by measuring the
frequency of radial breathing mode oscillations~\footnote{These measurements
are performed with sufficiently weak compensation that the potential remains
approximately harmonic.}. 
We find, up to a $\pm$5\% systematic uncertainty, the lattice beam waists to be 
$w_{L}=(47;47;44)\,\si{\micro\metre}$ and the compensation beam waists to be 
$w_{C}=(42;41;40)\,\si{\micro\metre}$.

\FigureTwo

We measure the {\it in-situ} column density distribution of the atoms
using polarization phase-contrast imaging~\cite{Bradley1997}.  This technique
can be used to image dense clouds, in contrast to absorption imaging which is
limited to small optical densities due to saturation.   The imaging light was
detuned by -150~MHz from state $|\spup\rangle$ (-74~MHz from $|\spdn\rangle$),
keeping the phase shift across the cloud below $\pi/5$ to avoid significant
dispersive distortions of the image.  

Figure~2 shows  azimuthal averages of the column density and density profiles;
the latter are obtained from the former using the inverse Abel transform (which 
assumes spherical symmetry)~\cite{scikit-image,Note3}.
Profiles for three different values of $U_{0}/t_{0}$ (where $U_{0}$ and $t_{0}$ 
denote the values of the Hubbard parameters at the center of the trap) are 
shown, along with profiles calculated for our trap potential.   

For the numerical calculations, we set $T$ and the global chemical potential,
$\mu_{0}$, while the local values of $U/t$, $T/t$, and $\mu/t$ are calculated
using the known trap potential.  Local values of the density are obtained,
within the LDA, by interpolation of NLCE and DQMC results for a homogeneous
system calculated in a $(U/t,T/t,\mu/t)$ grid.  
Because $T/t$ diminishes with $r$, the lowest value of $T/t_{0}$ that can be
calculated for the trap is limited to $T/t_{0}=0.6$.

\FigureThree

The response of the central density of the cloud, $\tilde{n}_{0}$, to changes
in atom number,  is a measure of the local compressibility at the center of the
trap.  We obtain $\tilde{n}_{0}$ by fitting the measured column density with
the integral, $\int\tilde{n}(\rho,z)\,\mathrm{d}z$, of a flat-topped Gaussian
function 
\begin{equation}
 \tilde{n}(\rho,z) = \begin{cases} 
  \tilde{n}_{0}  ~~~\text{if\ \ \   $\rho^{2} + z^{2} < r_{0}^{2}  $}  & 
      \vspace{0.9em} \\   
  \tilde{n}_{0} \exp\left[ \frac{ r_{0}^{2} - \rho^{2} - z^{2}}
                        { \sigma^{2}  } \right]  ~~~ 
                    \text{otherwise} & 
  \end{cases}, 
\end{equation} 
where $\rho$ is the distance from the imaging axis, and the fit parameters are
$\tilde{n}_{0}$, the flat-top radius, $r_{0}$, and the Gaussian $1/e$ radius of
the cloud's wings, $\sigma$.  In Fig.~3 we show $\tilde{n}_{0}$ vs.~$N$ for
various values of the interaction strength $U_{0}/t_{0}$.  The appearance of a
plateau in $\tilde{n}_{0}$ around 1 is characteristic of the Mott insulating
regime.  The persistence of a Mott plateau at intermediate coupling,
$U_{0}/t_{0}=11.1$, indicates that the temperature is at or below the tunneling
energy, as shown by comparison with the numerical results.  A precise
temperature determination is prevented by the fact that the density and other
observables related to the charge degrees of freedom, are relatively
insensitive to temperature for $T<t$.

The local compressibility, $\tilde{\kappa}$, is obtained by taking a derivative
of the measured and calculated density profiles as 
\begin{equation}
\tilde{\kappa} =   
             \frac{ (3\pi^{2})^{2/3} }{ 2 } 
             \frac{ \partial \tilde{n}^{2/3} }{ \partial r } 
      \left(   \frac{ \partial (\mu/t) }{ \partial r }   \right)^{-1},
\end{equation}
where the spatial derivative of the local chemical potential depends only on
the trap parameters.  For the data, the azimuthal average of the column
density, and the inverse Abel transform are noisy at small radii, so, to avoid
excessive noise in the determination of the radial derivative of
$\tilde{n}^{2/3}$, we restrict our analysis to $r/a > 12$.  Figure~4 shows
$\tilde{\kappa}$ vs $\tilde{n}$ for the experimental data and for density
profiles calculated at different temperatures.  A decrease of the
compressibility near $\tilde{n}\approx1$, as expected for a Mott insulator, is
observed for $U_{0}/t_{0}=11.1$ and 14.5.  As with the central density, the
weak sensitivity of $\tilde{\kappa}$ to $T$ at lower temperatures prevents
us from making a precise temperature measurement. However, the comparison of
the data with the numerical calculations at $T/t_{0}=0.6$, in both Figs.~3 and
4, reveals that the results are consistent with our previous measurement in the
same system, where using spin-sensitive Bragg scattering of light, we
determined the temperature to be
$T/t_{0}=0.58\pm0.07$~\cite{Corcovilos2010,Hart2014arxiv,Note4}.

\FigureFour

We have shown that the local compressibility of a two-component Fermi gas in an
optical lattice may be extracted from {\it in-situ} measurements of the column
density.   The data presented here shows evidence of Mott-insulating behavior
for interaction strengths as low as $U_{0}/t_{0} = 11$,  close to where $T_{N}$
is expected to be maximal, and where AFM correlations were observed to be
maximal for this system~\cite{Hart2014arxiv}.  A key achievement of this work
is the combination of experiment with two complimentary theoretical approaches
which span the full range of $U/t$ and $\tilde{n}$ required to model the
trapped atom data.  As described in the supplemental material~\cite{Note5},
the use of DQMC and NLCE in tandem provides reliable
results over a range of temperatures and interaction strengths beyond those
available previously.
{\color{white} \cite{Fuchs2011,Rohringer2011,Henderson1992}}

Measurements of local compressibility in an optical lattice, along with
recently developed methods for detecting magnetic order, can improve our
understanding of the onset of Mott insulating behavior in the Hubbard model,
and answer open questions about its proximity to the AFM phase in different
coupling regimes. In addition, the local compressibility can have important
implications for understanding the nature and extent of the non-Fermi liquid
state of the 2D Hubbard model away from
half-filling~\cite{PhysRevLett.63.1996,PhysRevLett.102.206407,
sordi2012pseudogap} at relatively high temperatures~\cite{PhysRevB.80.140505}.
Finally, as has been recently
shown~\cite{PhysRevB.81.201101,PhysRevB.80.245102}, sharp signatures of phase
separation and stripe formation are evident in the compressibility, raising the
possibility that this central property of cuprate superconductors, and of the
Hubbard model, might be accessible to this diagnostic.

\begin{acknowledgments}
This work was supported under ARO Grant No.\,W911NF-13-1-0018 with funds from
the DARPA OLE program, NSF, ONR, the Welch Foundation (Grant No. C-1133), and
ARO-MURI Grant No. W911NF-14-1-0003. T.P. acknowledges support from CNPq,
FAPERJ, and the INCT on Quantum Information. N.T. acknowledges support from
grant No. NSF-DMR1309461.  R.T.S. acknowledges support from the University of
California, Office of the President.  T.P. and R.T.S. acknowledge funding from
Science Without Borders, Brazil.
\end{acknowledgments}

%

\pagebreak
\onecolumngrid
\begin{center}
\textbf{\large Compressibility of a fermionic Mott insulator of ultracold atoms:  Supplemental Material}
\end{center}
\twocolumngrid
\setcounter{equation}{0}
\setcounter{figure}{0}
\setcounter{table}{0}
\setcounter{page}{1}
\makeatletter
\renewcommand{\theequation}{S\arabic{equation}}
\renewcommand{\thefigure}{S\arabic{figure}}
\renewcommand{\bibnumfmt}[1]{[S#1]}
\renewcommand{\citenumfont}[1]{S#1}


For the numerical calculations used to benchmark the experimental data in
the paper, we used the local density approximation (LDA) and  a combination of
determinantal quantum Monte Carlo (DQMC) and numerical linked-cluster expansion
(NLCE) results for a homogeneous system,  calculated in a $(U/t, T/t, \mu/t)$
grid.  The use of DQMC and NLCE in tandem provides results over a range of
temperatures and interaction strengths beyond those available previously,
including using DQMC alone~\cite{Paiva2011S}, the dynamical cluster
approximation (DCA)~\cite{Fuchs2011S}, diagrammatic QMC~\cite{Kozik2013S}
or the dynamical vertex approximation~\cite{Rohringer2011S}. 

Previous work coupled experimental values for the global compressibility with
theoretical calculations in the atomic limit~\cite{Jordens2008S}, with dynamic
mean field theory (DMFT)~\cite{Schneider2008S}, or with a high-temperature
series expansion (HTSE)~\cite{Jordens2010S,Taie2012S}.  These approaches capture
the qualitative physics of the Mott transition, but ultimately become
inaccurate as the temperature decreases, limiting their usefulness as
experimental benchmarks.  

\begin{figure}[h]
\centering\includegraphics[width=0.62\columnwidth]{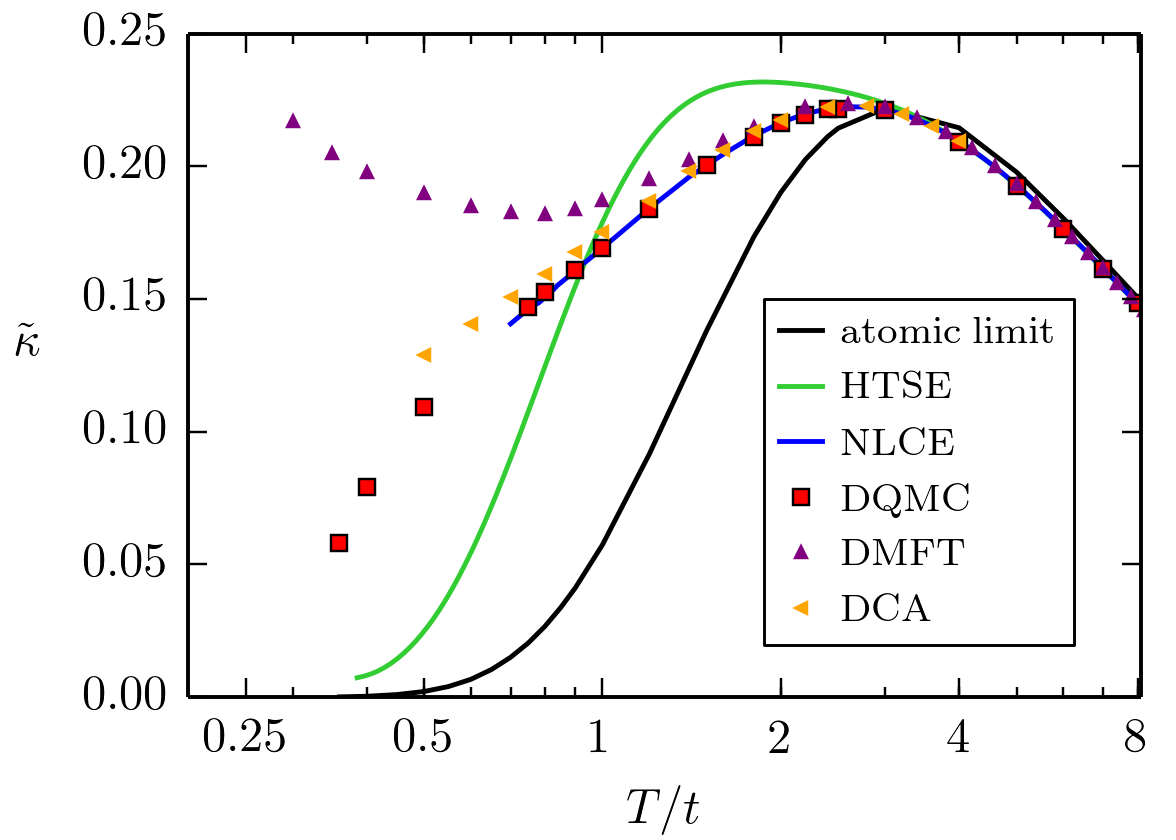}
\caption{Normalized compressibility, $\tilde{\kappa}$, at $\tilde{n}=1$
(half-filling) and $U/t=8$ versus $T/t$.  The atomic limit is the limit where
$t\rightarrow 0$;  the HTSE is an expansion to second order in
$t/T$~\cite{Henderson1992S}; the NLCE is carried out to the eighth order in the
site expansion~\cite{Khatami2011S}; DQMC is calculated in a
$6\!\times\!6\!\times\!6$ lattice, according to the methodologies in
Refs.~\cite{Blankenbecler1981S,Paiva2010S};  DMFT data was obtained from
J.~Imri\v{s}ka and T.~Sch\"{a}fer~\cite{Note6}; DCA data, calculated as
outlined in Ref.~\cite{Fuchs2011S}, was obtained from E.~Gull~\cite{Note7}.}
\label{fig:one}
\end{figure}

In the case of the atomic limit, deviations (at half-filling) from more refined
treatments like DQMC and NLCE, used in this paper, or the DCA~\cite{Fuchs2011S}
begin at $T/t \sim 2$, as shown in Fig.~\ref{fig:one} for $U/t=8$.  DMFT is
accurate to lower $T$, but is also known to exhibit low $T$ pathologies, most
notably over-estimating the anomalous increase in double occupancy as $T$ is
lowered~\cite{Paiva2011S,Fuchs2011S}.   A comparison at $U/t=4$ revealed that the
temperature at which the different theories deviate from each other is similar
to that seen at $U/t=8$.

We have focused our comparison of different theoretical methodologies on
half-filling ($\tilde{n}=1$) and an intermediate interaction strength, $U/t=8$,
since these are optimal parameters to observe antiferromagnetic correlations,
and hence much experimental attention is focused on maximizing the fraction of
the confined cloud at these conditions.  The evolution of the accuracy of these
approaches with $U/t$ is expected to be more complex.  The atomic limit, the
HTSE,  and the NLCE are particularly suited to large $U/t$, and indeed the
latter is the method of choice for $U/t \gtrsim 10$ and temperatures in the
ranges thus far accessible to experiment.  At weak $U/t$, DMFT results will,
among other things, be affected by the assumed form of the non-interacting
density of states (DOS), e.g.  choosing the semicircular DOS versus the
tight-binding model on the cubic lattice. These have different values of the
second moment, which introduces modest differences in $\tilde{\kappa}$ at low
$T$.

\onecolumngrid

\twocolumngrid

\underbar{Acknowledgments:}
We are very grateful to Emanuel Gull, Jacub Imri\v{s}{}ka, and Thomas
Sch\"afer for discussions and data (DMFT and DCA).

%

 
\end{document}